\documentstyle{article}
\def\slashmark#1#2#3{\global\setbox0=\hbox{\raise#2em
	\hbox{\kern#3em $#1\mathchar"0236$}}%
	\wd0=0pt \ht0=0pt \dp0=0pt \box0}

\def\slash#1{\raise.18ex\hbox{/}\kern-.68em #1}
\begin{document} \newcommand{\beq}{\begin{equation}}
\newcommand{\eeq}{\end{equation}} \newcommand{\bea}{\begin{eqnarray}}
\newcommand{\eea}{\end{eqnarray}} \newcommand{\nn}{\nonumber}
\newcommand{\bal}{\begin{array}{ll}} \newcommand{\eal}{\end{array}}
\def\1{{\rm 1 \kern -.10cm I \kern .14cm}} \def\R{{\rm R \kern -.28cm I
\kern .19cm}}

\begin{titlepage}
\begin{flushright}  LPTHE Orsay 94/73 \\SPhT Saclay T94/145 \\November 1994
\end{flushright}
\vskip 1cm
\centerline{\LARGE{\bf {Fermion mass
hierarchies in low energy}}}
\centerline{{\LARGE{\bf {supergravity and
superstring models}}}\footnote{Supported in part by the CEC Science
project no. SC1-CT91-0729.}}
\vskip 2cm \centerline{\bf {Pierre Bin\'etruy}}  \vskip .5cm
\centerline{\em Laboratoire de Physique Th\'eorique et Hautes Energies
\footnote{Laboratoire associ\'e au Centre National de la Recherche
Scientifique }}
\centerline{\em
 Universit\'e Paris-Sud, B\^at.
211, F-91405 Orsay Cedex, France }
\vskip 1cm \centerline{\bf {Emilian Dudas}} \vskip .5cm
\centerline{\em Service de Physique Th\'eorique de Saclay,}
\centerline{\em
91191 Gif-sur-Yvette Cedex, France}
\vskip 1cm
\centerline{\bf {Abstract}}

\indent

We investigate the problem of the fermion mass hierarchy in supergravity models
with flat directions of the scalar potential
associated with  some gauge singlet moduli fields.
The low-energy Yukawa couplings are nontrivial homogeneous functions
of the moduli and a geometric constraint between them plays, in a large
class of models, a crucial role in generating hierarchies. Explicit
examples are given for no-scale type supergravity models.
The Yukawa couplings are dynamical variables at
low energy, to be determined by a minimization process which amounts
to fixing ratios of the moduli fields. The Minimal
Supersymmetric Standard Model (MSSM) is studied and the constraints
needed on the parameters in order to have a  top quark much heavier
than the other fermions  are worked out. The bottom mass is explicitly
computed and shown to be compatible with the experimental data for a
large region of the parameter space.

\end{titlepage}

\section{Introduction} \label{sec:nom1}

One of the mysteries of the Standard Model of strong and electroweak
interactions
is the difference between the mass of the top quark and the mass of the
other fermions. Taking as fundamental  the electroweak scale $v
\simeq 250 GeV$, the top quark mass is roughly of the order $v$, whereas
in a first approximation all the other fermions are massless. The
Standard Model by itself cannot explain this puzzle ; by definition the
Yukawa couplings (Yukawas) are just free parameters, and so are the
physical fermion masses. Going beyond the Standard Model, Grand
Unified Theories and (or) Supersymmetry give some relations between the
Yukawas, but do not answer the question. The couplings are still free
parameters, to be eventually determined in a more fundamental theory.

An interesting idea was recently proposed in the context of the Standard
Model by Nambu \cite{Nam}.  Essentially the vacuum energy density is
minimized with respect to the Yukawa couplings $\lambda_i$, all the
other parameter being held fixed, including the vev's of the scalar
fields.  The Yukawas $\lambda_i$ are subject to a constraint, of
vanishing quadratic divergences in the Higgs sector of the theory.

This gives the so-called Veltman condition \cite {Vel}, which in the
case of two Yukawas $\lambda_1$ and $\lambda_2$ gives
\bea
 \lambda_1^2 +
\lambda_2^2 = a^2 , \label{eq:ar1}
\eea
 where a is a constant. The
vacuum energy to be minimized in the example chosen by Nambu is of the
form
\bea
{\cal E}_0 = - A \ (\lambda^4_1 + \lambda^4_2 ) + B \ (\lambda^2_1
\ln \lambda^2_1 + \lambda^2_2 \ln \lambda^2_2 ) \ . \label{eq:ar2}
\eea

Supposing for the moment that $B=0$ and minimizing ${\cal E}_0$, the
minimum is obtained for $(\lambda^2_1 , \lambda^2_2) = (a^2,0)$ or
$(\lambda^2_1 , \lambda^2_2) = (0,a^2)$. The configuration $(\lambda^2_1
, \lambda^2_2)  = ({a^2 \over 2} , {a^2 \over 2})$ it is a local
\underline{maximum} for ${\cal E}_0$ and it is the only extremum of
${\cal E}_0$.

In this approximation $(B=0)$ we have a massless fermion and a massive
one, with a mass fixed by the mass parameter of the Lagrangian. An
important constant is the sign of $A$, which should be positive ; this
will be an important argument in favor of supersymmetric theories later
on. Adding the logarithmic terms $B\not = 0$ has as effect producing a
global minimum in ${\cal E}_0$ , and the corresponding configuration is
$(\lambda^2_1 , \lambda^2_2) \simeq (a^2 , a^2 e^{-{2a^2 A \over B}})$
and the similar one $\lambda_1 \leftrightarrow \lambda_2$. The ratio of
the two fermion masses contains an exponential suppression factor
$e^{-{a^2A \over B}}$, and the hierarchy is obtained if $B <<a^2 A$.

The applicability of the mechanism to the standard model is under
investigation \cite{Gher}.

In a previous paper \cite{BD} we argued that a natural framework to
incorporate the Nambu idea is provided by the string effective
supergravity models. In this case the spectrum of the theory includes
gauge singlets $T_i$ called moduli which describe the size and the shape
of the six-dimensional compactified manifold. Their vev's are not
determined at the supergravity level and the scalar potential has flat
directions due to some non compact symmetries.

The coupling constants at low-energy are functions of the moduli and
consequently can be considered as dynamical variables to be determined
by the low-energy dynamics, in addition to the gravitino mass $m_{3/2}$.

This is possible because the breaking of supersymmetry destroys in
principle the non-compact symmetries and dynamically determines the
moduli vev's. The minimization with respect to the gravitino mass
$m_{3/2}$ was extensively studied in the literature \cite{ELNT} in
the content of the supergravity no-scale models \cite{CFKN} with one
modulus field $T$. This
case corresponds to the simplest case of Calabi-Yau type
compactification \cite{Witten} of the ten-dimensional heterotic string
theory, but is not realistic because it only allows to obtain  one
generation of fermions at low energy. In more realistic models
\cite{CHSW} we normally have several moduli and consequently more dynamical
variables. The low-energy determination of the Yukawas is equivalent to
the determination of the real part of the moduli fields. This hides in
fact an important hyphothesis, the masses of the moduli should be very
small compared to the intermediate supersymmetry breaking scale.
This is a nontrivial assumption and is a direct generalization of the
no-scale idea developed in \cite{ELNT}. If the moduli masses are much
greater than the electroweak scale, the moduli decouple at low-energy
and the Yukawa couplings are just arbitrary, fixed parameters and not
dynamical variables.

The purpose of the present paper is to present explicit
examples of supergravity theories which illustrates the previous ideas.
The characteristic feature of the models described below is the
particular way of realizing a constraint of type (\ref{eq:ar1}) between
Yukawas, directly at the tree level of supergravity .
The constraint, independent of the mechanism of supersymmetry breaking,
 is valid at the Planck scale and should be run to low energy using the
renormalization group (RG) equations.
Another possibility, related to the moduli dependent threshold corrections in
the string effective
supergravities will not be discussed here. More models will be
proposed with different constraints and the phenomenological
consequences for the low-energy fermion spectrum will be analyzed. As a
result, we will find that in most of the cases the spectrum consists
of one massive fermion, all the other being massless in a certain
approximation which corresponds to the case $B=0$ in the toy model
(\ref{eq:ar2}) discussed by Nambu.

The article is organized as follows. In section \ref{sec:nom2} we
analyze different supergravity models and the resulting constraints
between the low-energy Yukawas. The simplest models proposed give
multiplicative type constraints instead of the additive one in
eq.(\ref{eq:ar1}).
Models with additive constraints are somewhat more complicated and
possess less symmetries.

In section\ \ref{sec:nom3} the different constraints will be analyzed
in connexion with the minimization of the vacuum energy minimization.
It is shown that for
multiplicative type constraints in models with Higgs fields of the type
MSSM, a condition on the dilaton field $S$ is necessary in order to
generate hierarchies. In an example studied in section (3) this is
$(s+s^+)^2 > {4\pi^2 / \ell n{M_p \over \mu_0}}$, where $\mu_0$ is the
low-energy scale which sets the mass value for the massive fermion and
$M_p$ is the Planck scale.

Section\ \ref{sec:nom4}
discusses the Minimal Supersymmetric Standard Model \cite{Fay} and the
phenomenological constraints necessary for the mechanism to work.
It is shown that a minimal value for $tg \beta$ of order one is
sufficient to trigger the mechanism in such a way that an up-type
quark be the heaviest one. Keeping only the top and bottom Yukawa
couplings, we compute analytically the latter as a function of the
gauge coupling at the Planck scale and of the low energy parameters. We
find that it is directly proportional to the $\mu$-parameter of the MSSM.
For $tg \beta > 1$, we find that the bottom mass is compatible with
the experimental data for a large allowed region of the parameter space.

Finally some conclusions are drawn and prospects for future work are
presented.

\section {Dynamical Yukawa couplings and constraints in low-energy
supergravity models.}\label{sec:nom2}

In the following, we will only consider models with zero cosmological
constant at the tree level. The gauge singlet fields considered in all
examples will be the moduli $T_i$ and a dilaton-like field $S$, common
to all superstring effective supergravities. We will consider $N=1$
supergravity described by the K\"ahler function $K$, the superpotential
$W$ and the gauge kinetic function $f$\cite{CFGP}. We are not
interested in this section in the gauge interactions and consequently
we will neglect them in the analysis; we will return  to this when
discussing the MSSM.

Consider a string effective model containing the above-mentioned
singlet fields and
$p$-observable chiral fields $\phi^i_A$. The K\"ahler potential and the
superpotential read
\bea
 \bal K &=\ K_0 + {K_A}^{j_A}_{i_A} \phi^{i_A}
\phi^+_{j_A} + \cdots \ , \\ \\ K_0 &=\ -{3 \over n} \
\sum_{\alpha = 1}^n \ln (T_\alpha + T_\alpha^+ ) - \ln (S + S^+)
\\ \\ W &= \ {1 \over 3}\ \lambda_{i_A i_B
i_C} \phi^{i_A}
\phi^{i_B} \phi^{i_C} \ , \eal \label{eq:ar3}
\eea
where the dots stand for higher-order terms in the fields $\phi^{i_A}$.
The index $A$ in eq.(\ref{eq:ar3}) stands for sectors of the matter
fields with different modular weights \cite{DHVW}.
The K\"ahler metric depends on the moduli $T_{\alpha}$, and
eventually on $S$. The
low-energy spontaneously broken theory contains the normalized fields
$\hat \phi^i$ defined by $\phi^{i_A} = ({K_A}^{-1/2})_{j_A}^{i_A} \ \hat
\phi^{j_A}$ and the
Yukawas $\hat \lambda_{i_A i_B i_C}$. In order to obtain the relation between
$\lambda_{i_A i_B i_C}$ and $\hat \lambda_{i_A i_B i_C}$ , consider the
scalar potential
\cite{CFKN}, which contains the piece
\bea
 V \ = \ e^K \left(\sum_{A} ({K_A}^{-1})_{j_A} ^{i_A}\
D_{i_A} W \bar D^{j_A}\ \bar W - 3 |W|^2 \right) \ni
\hat W_{i_A}\ \hat{\bar W}^{i_A} \ . \label{eq:ar4}
\eea
 In eq.(\ref{eq:ar4}), $D_{i_A} = {\partial W / \partial \phi^{i_A}} + K_{i_A}
W$ and $\hat W = {1 \over 3} \hat \lambda_{i_A i_B i_C} \hat \phi^{i_A} \hat
\phi^{i_B} \hat \phi^{i_C}$ is the low-energy superpotential: we restrict
our attention here to the trilinear (renormalisable) couplings.
Making the identifications in eq.(\ref{eq:ar4}) we obtain the relation
\bea
 \hat
\lambda_{i_A i_B i_C} = e^{i \theta_{i_A}} \ e^{K_0 \over 2}
({K_A}^{-1/2})_{i_A}^{j_A} ({K_B}^{-1/2})_{i_B}^{j_B}
({K_C}^{-1/2})_{i_C}^{j_C} \lambda_{j_A j_B j_C} \ , \label{eq:ar5}
\eea
where $\theta_{i_A}$ is an arbitrary real function of the moduli.
{}From (\ref{eq:ar5}) we see that the low-energy Yukawas $\lambda_{i_A i_B
i_C}$
are functions of the moduli through the Kahler potential $K$. Some of
the moduli are fixed to their vacuum energy values at high energies.
Others may still remain undetermined at low energies and correspond to
flat directions of the scalar potential. In this case, the low energy
Yukawa couplings $\hat{\lambda}_{i_A i_B i_C}$ are dynamical degrees of freedom
whose precise value may be fixed by the dynamics at low energies.

Consider a model with $M$ Yukawa couplings.
As we will see in the next two sections, in order to understand
dynamically the fermion mass hierarchy we are interested in models where
the low energy Yukawas are not independent but subject to a certain
number $p$ of constraints. Such constraints can be written in terms of
$p$ independent functions $F_i$, $i = 1...p$, such that they read
\bea
F_i (\hat \lambda_1 (T_{\alpha}), \cdots , \lambda_M (T_{\alpha}) ) =
C_i \ , \label{eq:ar500}
\eea
where $C_i$ are constants which do not depend on the moduli.
Differentiating (\ref{eq:ar500}) with respect to the moduli, we find $p$
systems of $n$ linear equations of $M$ variables (${\partial F_i \over
\partial \hat{\lambda}_I}$)
\bea
\sum_{I=1}^M
{\partial \hat \lambda_I \over \partial T_{\alpha}}
{\partial F_i \over \partial \hat \lambda_I} = 0 \label{eq:ar501}
\eea
which can easily be put in a matrix form. The condition to have $p$
independent eigenvectors for this matrix equation is
\bea
rank \left({\partial \hat \lambda_I \over \partial T_{\alpha}}\right)
  = min (M,n) - p
\ . \label{eq:ar91}
\eea

Consider for the moment the case where the number of the Yukawas is
equal to the number of moduli $n = M$. A low-energy
constraint between Yukawas can be expressed mathematically in the
following way. Eliminating the moduli fields $T_i$ as functions of
the Yukawas, the transformation is singular
\bea
 \det ({\partial \hat
\lambda_I \over \partial T_\alpha} )  = 0 \ .
\label{eq:ar6}
\eea
The Jacobian (\ref{eq:ar6}) can be rewritten as follows
\bea
 \left | \begin{array}{c} \sum_\alpha T_\alpha
{\partial \hat \lambda_1 \over \partial T_\alpha } \;\; {\partial \hat
\lambda_1 \over \partial T_2 } \;\;\cdots\;\;
 {\partial \hat \lambda_1 \over \partial T_n } \\ \\ \sum_\alpha
T_\alpha {\partial \hat \lambda_M \over \partial T_\alpha } \;\; {\partial
\hat \lambda_M \over \partial T_2 } \;\;\cdots \;\; {\partial
\hat \lambda_M \over \partial T_n } \end{array} \right | = 0 \ .
\label{eq:ar7}
\eea
 A natural solution for (\ref{eq:ar7}) is $\sum_\alpha T_\alpha
{\partial \hat \lambda_I \over \partial T_\alpha} = 0$, in which case
the Yukawas $\hat \lambda_{i_A i_B i_C}$ are {\it homogeneous functions
of the moduli}. In other words, in any model with an equal number
of Yukawas and moduli, if all the Yukawas are non trivial homogeneous
functions, we will always have a constraint between them. This is so
for many effective string models and will be our main assumption in
the following.\footnote{Similarly,
the condition to have $p$ constraints in the
general case $n \not=M$ is easy to find. We must consider all the
quadratic matrices $[min(M,n)] \times [min(M,n)]$ constructed from
the matrix $[\partial \hat
\lambda_I / \partial T_\alpha]$ and impose the condition that the
rank of all of them  be $min(M,n) - p$. As above, this is so if any
subset of $min(M,n) - p + 1$
Yukawa couplings obey homogeneity properties
with respect to any subset of $min(M,n) - p + 1$ moduli.}

Define the modular weights of the matter fields as
\bea
T_\alpha {\partial \over \partial T_\alpha} {K_A}_{i_A}^{j_A} = {n_A}
{K_A}_{i_A}^{j_A} \ .
\label{eq:ar8}
\eea
Using eq.(\ref{eq:ar5}), the homogeneity property of
$\hat \lambda_{i_A i_B i_C}$ translates
into an equation for the original couplings $\lambda_{i_A i_B i_C}$
\bea
\left ( {1 \over 2} T_{\alpha} K^{\alpha} - {n_A + n_B + n_C \over 2} +
T_{\alpha} {\partial \over \partial T_{\alpha}} \right ) \lambda_{i_A
i_B i_C} = 0 \ . \label{eq:ar502}
\eea
For theories with zero cosmological constant at tree level
$T_{\alpha} K^{\alpha} = -3$.
Denoting by $N_{ABC}$ the modular weight of the string couplings
$\lambda_{i_A i_B i_C}$, eq.(\ref{eq:ar502}) reduces to
\bea
n_A + n_B + n_C = -3 + 2 N_{ABC} \ . \label{eq:ar 503}
\eea
This equation is sufficient to guarantee the existence of constraints
between Yukawa couplings. Since it depends only on the modular weights
of the fields and their couplings, it proves to be useful to construct
explicit models.

A simple case of interest is $N_{ABC} = 0$ and ${K_A}_{i_A}^{j_A} =
{c_A} {t_A}^{n_A} {\delta}_{i_A}^{j_A}$ where $c_A$ are constants and
$t_A = T_A + T^+_A$.
The homogeneity property becomes explicit and the equation
(\ref{eq:ar5}) becomes
\bea
{\hat \lambda}_{i_A i_B i_C} = (s c_A c_B c_C)^{-{1 \over 2}}
{\prod_{\alpha} ({t_{\alpha} \over t_C})^{-{3 \over 2n}}}
({t_A \over t_C})^{-{n_A \over 2}}
({t_B \over t_C})^{-{n_B \over 2}} \lambda_{i_A i_B i_C} \ ,
\label{eq:ar503}
\eea
where $s = S + S^+$.
A  very particular case is $N_{ABC} = 0$ and $n_A = n_B = n_C = -1$, in
which case we include the matter fields
$\phi^{i_A}$ in the no-scale structure of the moduli. The simple models
analyzed below will have this property which is typical of many
compactifications of the ten-dimensional heterotic string theory.

We can easily compare the number of degrees of freedom at low energy $(M + 1 )$
(one degree of freedom is the gravitino mass) and high-energy $(n)$ (we
consider only the real part of the moduli; the imaginary
part will play no role in the determination of the Yukawas in our
examples ). In order to completely fix the moduli vev's we
must satisfy the inequality $M+1 \geq n$. For $M=n$ the vev's are fixed
and moreover we have one constraint between the Yukawas. This is the
most interesting situation which, as emphasized above, will be our main
concern.

The symmetries of the supergravity models will be essential in order to
restrict the class of possible constraints. An important invariance is
provided by the K\"ahler transformations
\bea
 \bal K(z,z^+) &\rightarrow
K(z,z^+ ) + F(z) + F^+(z^+) \ , \\ \\ \hskip 1cm W(z) &\rightarrow
e^{-F(z)} W(z) \ , \eal \label{eq:ar9}
\eea
 where $F(z)$ is an analytic
function of the set of chiral superfields $z$. If we restrict our
attention to moduli dependent functions, this transformation acts
on the low energy Yukawas as a $U(1)$ transformation :
\bea
 \hat
\lambda_{i_A i_B i_C} \rightarrow e^{-i Im F} \hat \lambda_{i_A i_B i_C} . \
\label{eq:ar10}\eea
 The transformation (\ref{eq:ar9}) allows us to eliminate the phase
$\theta_{i_A}$ in eq.(\ref{eq:ar5}) and tells us that the constraint
must always contain
the combination $\hat \lambda \hat \lambda^+$. This is not very
restrictive but it ensures that our final results are K\"ahler
invariant.

A  more powerful constraint  is obtained if we impose the
target-space duality symmetries $SL(2,Z)$
\bea
 T_{\alpha}
\rightarrow {a_{\alpha} T_{\alpha} - i b_{\alpha} \over i c_{\alpha}
T_{\alpha} + d_{\alpha}}\ ,\ a_\alpha d_\alpha - b_\alpha c_\alpha = 1,
\ a_\alpha , \cdots , d_\alpha \in Z  \ , \label{eq:ar11}
\eea
 for every moduli $T_\alpha$. Since
\bea
 T_\alpha + T^+_\alpha
\rightarrow {T_\alpha + T^+_\alpha \over | i c_\alpha T_\alpha +
d_\alpha |^2} \ , \label{eq:ar12}
\eea
 it can  be viewed as a
particular type of K\"ahler transformations, acting explicitly on the
fields $\phi^{i_A}$.  In  effective string
theories of the orbifold type \cite{DHVW}, the observable fields
$\phi^{i_A}$ and the K\"ahler metric ${K_A}^{j_A}_{i_A}$
transform under (\ref{eq:ar11}) as
\bea
 \left \{ \bal
\phi^{i_A} &\rightarrow {\phi^{i_A} \ / ( i c_{\alpha_A} T_{\alpha_A} +
d_{\alpha_A})^{n_A}}\\ &\\ {K_A}_{i_A}^{j_A} &\rightarrow | i
c_{\alpha_A} T_{\alpha_A} +
d_{\alpha_A} |^{2 n_A} \ {K_A}_{i_A}^{j_A} \ , \label{eq:ar13} \eal \right.
\eea
 where
$T_{\alpha_A}$ is the moduli containing $\phi^{i_A}$ in its no-scale
structure (explicit examples will be given below). Hence the low-energy
fields $\hat \phi^{i_A} = ({{K_A}^{1/2}})_{i_B}^{i_A} \  \phi^{i_B}$ are
duality invariant.
For the model defined
by eq.(\ref{eq:ar3}) , the K\"ahler potential $K$ transforms as follows

\bea
 K \rightarrow K + {3 \over n} \sum_{\alpha_A = 1} ^n \ln  |i
c_{\alpha_A} T_{\alpha_A} + d_{\alpha_A} |^2 \ .  \label {eq:ar14}
\eea
 Defining the
function $F_\alpha \equiv {3 \over n} \ln (i c_\alpha T_\alpha +
d_\alpha)$, the Yukawa transformation law is obtained from eq.
(\ref{eq:ar5}). Assuming that the original Yukawa couplings
$\lambda_{i_A i_B i_C}$ are modular invariant ($N_{ABC}=0$),
one finds for the low energy couplings
\bea
 \hat \lambda_{i_A i_B i_C} \rightarrow
\bigl ( \prod_{\alpha_A =1}^n e^{{F_{\alpha_A} + F^+_{\alpha_A} \over 2}} \bigr
)
\ e ^{-{n \over 6}(F_{\alpha _A} n_A + F_{\alpha _B} n_B + F_{\alpha_C}
n_C  + h.c.
)} \ \hat \lambda_{i_A i_B i_C} \ . \label{eq:ar15}
\eea
We can easily see that a theory which is completely duality invariant is too
restrictive for us. The correct transformation of the superpotential
$W(z)$ in eq.(\ref{eq:ar9}) gives  the equality
\bea
\sum_{\alpha_A =1}^n {F_{\alpha_A}} = {n \over 3}(F_{\alpha _A} n_A
 + F_{\alpha _B} n_B  + F_{\alpha_C} n_C
) \ . \label{eq:ar100}
\eea
This tells us that in this
case the Yukawa couplings $\hat \lambda_{i_A i_B i_C}$ are duality
invariant, {\em i.e.} invariant under (\ref{eq:ar15})
and at the tree level of supergravity with no threshold corrections this
means that they do not depend on the moduli at all.

The most important
ingredient in the construction of realistic supergravity theories is
the geometrical structure of the Kahler potential, tranforming like
(\ref{eq:ar9}). The scalar fields  span homogeneous spaces of the
coset type, as in the examples presented below. We will consider
superpotentials which present only a part of the symmetries of the Kahler
function $K$, which are sufficient to guarantee the existence of the
flat directions and a zero cosmological constant.
This happens for example in the four-dimensional $N=1$ string
constructions with spontaneous supersymmetry breaking at the tree level
\cite{KP}. This mechanism can be formulated in the orbifold string
constructions and the superpotential modification associated with
supersymmetry breaking violates the target space duality \cite{FKPZ}.

We will use in this
sense the notion of duality invariant models in the rest of this paper.
Note from eq.(\ref{eq:ar5}) that the superpotential  does not appear
explicitly in the ratio $\hat \lambda_{i_A i_B i_C} /
\lambda_{i_A i_B i_C}$, giving a simple geometric interpretation
for this ratio.

An allowed constraint involves
a duality invariant combination of $\hat \lambda_{i_A i_B i_C}$. A
simple inspection of eq.(\ref{eq:ar15}) is sufficient to convince
ourselves that only multiplicative-type constraints can be duality
invariant, due to the exponential transformation law. For example, if
$\hat \lambda_{i_A i_B i_C} \not = 0$ for all $i_A , i_B , i_C$ we
get in an obvious way the constraint
\bea
 \prod_{i_A i_B i_C} \hat
\lambda_{i_A i_B i_C} = cst \ . \label{eq:ar16}
\eea
 In all the models
discussed below, however, we have some $\hat \lambda_{i_A i_B i_C} = 0$ and
eq.(\ref{eq:ar16}) does not apply directly.

The simplest example contains two moduli $T_1,T_2$, the dilaton $S$ and
two observable fields $\phi^i$. The model is defined by
\bea
 \bal K &=
-{3 \over 2} \ln (t_1 - |\phi_1|^2) - {3 \over 2} \ln (t_2 - |\phi_2
|^2 ) - \ln s \ , \\ &\\ W &= {1 \over 3} \lambda_1 \phi^3_1 +
{1 \over 3} \lambda_2 \phi^3_2 + W(S) \ , \eal \label{eq:ar17}
\eea
where $t_i = T_i + T^+_i, \ s = S + S^+$ and $W(S)$ is a
non-perturbative contribution to $W$ which
fixes the value of $S$ and simultaneously breaks supersymmetry, as in
the usual gaugino condensation scenario \cite{FGN}.

The tree-level scalar potential is given by the expression
\bea
 V_0 = {1 \over s (t_1 - |\phi_1 |^2 )^{3\over2} (t_2 -
|\phi_2 |^2 )^{3\over2}} \
\left \{ \left | S{\partial W / \partial S}- W \right | ^2 + {2 \over 3}
\sum_{i=1,2} (t_i - |\phi_i |^2 ) \left | {\partial W / \partial \phi_i
} \right | ^2 \right \}
 \label{eq:ar18}
\eea
 The minimum is reached for $\phi^i = 0$ , $S{\partial W / \partial S}
- W = 0$ and $T_i$ undetermined, with a zero
cosmological constant. $V_0$ has flat directions in the $(T_1 , T_2 )$
plane and the K\"ahler function parametrizes a $\left [ {SU(1,2) / {U(1)
\times SU(2)} } \right ] ^2 \times {SU(1,1) / U(1)}$ K\"ahler manifold.
The superpotential $W$ breaks the $[SU(1,2)]^2$ symmetry associated with
the moduli down to $U(1)^2 \times$ diagonal dilatation.
The residual symmetry is written explicitly in eq.(\ref{eq:ar34})
and is spontaneously broken together with  supersymmetry.
In the low energy limit $M_P \rightarrow \infty$, the supersymmetric
scalar potential reads
\bea
 V_0 = \hat \lambda_1^2 |\hat \phi_1 |^4 +
\hat \lambda_2^2 |\hat \phi_2 |^4 \ . \label {eq:ar19}
\eea
 The low energy
Yukawas as functions of the high-energy $\lambda_i$ read from eq.(\ref{eq:ar5})
\bea
 \hat
\lambda_1^2 &=& {8 \over 27} \ {1 \over s} \left ( {t_1 \over t_2}
\right ) ^{3/2} \lambda_1^2 \ , \nonumber \\ \hat \lambda_2^2 &=& {8
\over 27} \ {1 \over s} \left ( {t_2 \over t_1} \right ) ^{3/2}
\lambda_2^2 \ . \label{eq:ar20}
\eea
 They are homogeneous functions of
the moduli and, consequently, the Jacobian \ $\det \left ({\partial \hat
\lambda_i \over \partial t_\alpha } \right ) = 0$ as can be explicitely
verified in eq.(\ref{eq:ar20}). $\hat \lambda_i$ are dynamical variables
at low energy , together with the gravitino mass $m^2_{3/2} = |W|^2 /
(s t_1^{3/2} t_2 ^{3/2})$. The constraint between Yukawas is
obvious from eq.(\ref{eq:ar20})
\bea
 \hat \lambda_1 \hat \lambda_2 = {8
\over 27} \ {1 \over s} \ \lambda_1 \lambda_2 \equiv \widetilde a ^2 =
\mbox{fixed} \ . \label{eq:ar21}
\eea
This eliminates one of two Yukawas as a dynamical variable and leaves
us with two variables, corresponding to
the two original moduli $T_1,T_2$. The minimization process at
low energy will partially fix the vacuum state and lift
the flat directions corresponding to $Re \ T_1$ and $Re \ T_2$. We are still
left with  flat directions for the imaginary parts of the moduli,
$Im \ T_1$ and $Im \ T_2$.

Equation (\ref{eq:ar21}) is valid at the Planck scale $M_p$. In the
effective theory at lower scales $\mu$ we must use the renormalization
group (RG) equations in order to express it as a function of $\hat
\lambda_i(\mu)$. This analysis, the comparison with a Veltman-type
constraint and the phenomenological consequences of minimization will be
analyzed in the next section.

The importance of putting $\phi_1$ and $\phi_2$ in different no-scale
structures can be seen by considering a slight modification of
eq.(\ref{eq:ar17}), with the same superpotential $W$ and the K\"ahler
potential
\bea
K = -{3 \over 2} \ln (t_1 - |\phi_1|^2 - |\phi_2
|^2) - {3 \over 2} \ln (t_2) - \ln s \ . \
\eea
In this case the low energy couplings have the same dependence on the
moduli
\bea
 \hat
\lambda_1^2 &=& {8 \over 27} \ {1 \over s} \left ( {t_1 \over t_2}
\right ) ^{3/2} \lambda_1^2 \ , \nonumber \\ \hat \lambda_2^2 &=& {8
\over 27} \ {1 \over s} \left ( {t_1 \over t_2} \right ) ^{3/2}
\lambda_2^2
\eea
and the analog of the constraint (\ref{eq:ar21}) is now a
proportionality relation
\bea
{{\hat \lambda_1}^2 \over \lambda_1^2} = {{\hat \lambda_2}^2 \over
\lambda_2^2} \ . \
\eea
As we will see in the next section, this kind of proportionality does
not lead to any hierarchy of couplings. In what follows, we will
consequently put in {\it different} no-scale structures the
different quark-type fields among which we want to generate a
hierarchy.

The generalization to more couplings of the model in eq.(\ref{eq:ar17})
is straightforward. The theory is described by
\bea
 \bal K &= -{3 \over
n} \sum_{i=1}^n \ln (t_i - |\phi_i|^2) - \ln s \ , \\ &\\ W
&= {1 \over 3} \sum_i \lambda_i \phi^3_i \ .
 \eal \label{eq:ar22}
\eea
 The low-energy Yukawas are given by (from now
on we will take $\hat \lambda$ real)
\bea
 \hat \lambda_i^2 = \left ({n
\over 3} \right )^3 {t_i^3 \over s \prod_{j=1}^n t_j^{3 \over n}}
\lambda_i^2 \label{eq:ar23}
\eea
 and the resulting constraint is
\bea
\hat \lambda_1 \cdots \hat \lambda_n = \left ({n \over 3} \right )^{3n
\over 2} {\lambda_1 \cdots \lambda_n \over s^n} \ = \ \mbox{fixed}
\ . \label{eq:ar24}
\eea

The important point to retain is that different observable fields
$\phi^i$ are included in different no-scale structures, corresponding to
different moduli. Including two observable fields in the same moduli
structure will produce proportional Yukawas and not multiplicative
constraints between them. In the following section we will see that
multiplicative constraints are essential in this framework to
understand why one low-energy
fermion is much heavier than the other fermions.

More possibilities are left when one introduces scalar fields to play
the role of the Higgs fields of the MSSM. For example, a model defined
by

\bea
 \bal K &= -{3 \over n} \sum_{i=1}^n \ln (t_i - |\phi_i|^2\ -
|H_i|^2) - \ln s \ , \\ &\\ W &= {1 \over 2} \sum \lambda_i
\phi^2_i H_i \ , \eal \label{eq:ar25}
\eea
 with an observable field
$\phi^i$ and a Higgs $H_i$ corresponding to one moduli $T_i$ gives the
same constraint as the preceding model, eq.(\ref{eq:ar24}).

If we want to couple more observable fields to the same Higgs field, we
can consider the model
\bea
 \bal K &= -{3 \over n} \sum_{i=1}^{n-1} \
\ln (t_i - |\phi_i|^2) -{3 \over n} \ \ln (t_n - |H|^2 ) - \ln s \ , \\
&\\ W &= {1 \over 2} ( \lambda_1 \phi^2_1+\cdots + \lambda_{n-1}
\phi^2_{n-1} ) H + {1 \over 3} \lambda_n H^3 \ . \eal \label{eq:ar26}
\eea

We introduced a special modulus for the Higgs H in order not to break the
permutation symmetry between the observable fields and to keep at
the same time the duality symmetries. The constraint is easily
obtained by defining ratios of
moduli of the type $A_i = {t_i / t_1} , i > 1$. In this way the
low-energy Yukawas depend only on $(n - 1)$ variables ,
\bea
 \left \{
\bal \hat \lambda^2_1 = \left({n \over 3}\right)^3 {1 \over s}
{A_n \over (A_2 \cdots A_n )^{3 \over n} } \
\lambda^2_1 \\ \\ \hat \lambda^2_i = \left({n \over 3}\right)^3
{1 \over s} {A^2_i A_n \over (A_2 \cdots A_n)^{3 \over n}
}\ \lambda^2_i \ , \ i = 2, \cdots , n \ .  \eal \right .
\label{eq:ar27}
\eea
 Eliminating $A_i = (\lambda_1 / \lambda_i) (\hat
\lambda_i / \hat \lambda_1)$, we get the constraint
\bea
 \hat \lambda_1
\cdots \hat \lambda_{n-1} = C \ \hat \lambda^{{n \over
3}-1}_n \ .  \label{eq:ar28}
\eea
where $C = \left({n \over 3}\right)^n \left({1 \over s}\right)^{n/3}
\lambda_1 \cdots \lambda_{n-1} \lambda_n^{1-n/3}$.
This constraint is multiplicative and symmetric in
the Yukawas $\hat \lambda_1 ... \hat \lambda_{n-1} $ (asymmetric
constraints  are easy to obtain from asymmetries in the K\"ahler
potential) and only $\hat
\lambda_n$ plays a particular role. The multiplicative constraints
satisfy automatically the K\"ahler invariance, eq.(\ref{eq:ar9}). In all
our examples this will be automatic, because the models are K\"ahler
invariant (in the specific sense described before) at  tree level.

We can consider a model which is the closest one to the minimal
non-minimal extension of MSSM \cite{NSW}, described by
\bea
 \left \{
\bal W = {1 \over 2} \sum_i (\lambda_i \phi_i^2 ) H_1 + {1 \over 2}
\sum_\alpha (\lambda_\alpha \phi^2_\alpha ) H_2 + \lambda_Y H_1 H_2 Y +
{k \over 3} Y^3 \ , \\ \\ K = -{3 \over n+2} \sum_{i=1}^{n_1} \ell n(t_i
- | \phi_i |^2 ) - {3 \over n+2} \sum_{\alpha =1}^{n_2} \ell n (t_\alpha
- | \phi_\alpha |^2 )  \\ \\ \hskip 1,3cm -{3 \over n+2} \ell n (t_{n+1}
- |H_1 |^2 - |H_2 |^2) - {3 \over n+2} \ell n(t_{n + 2} - |Y|^2 ) \ ,
\eal \right . \label{eq:ar31}
\eea
 with $M_1 + M_2 = n$. As all the
previous examples, this model has a duality invariance with respect to
all moduli $T_i$ and the number of moduli is equal to the number of
Yukawas. The two Higgs are put in the same moduli structure $T_{n+1}$,
but as before any observable quark field $\phi^i$ is associated with a
different modulus. The constraint is computed in the same way as in the
model of eqs.(\ref{eq:ar26}, \ref{eq:ar28}). The result is
\bea
 \hat
\lambda_1 \cdots \hat \lambda_n \ = \ \mbox{fixed} \times \ {\hat
\lambda_Y^{{n-2 \over 2}} \over \hat k^{n+2 \over 6} } \label{eq:ar32}
\eea
 and is symmetric in $\hat \lambda_1 , \cdots , \hat \lambda_n$.

All the examples discussed above are simple and have the duality
invariances, eq.(\ref{eq:ar11}). The symmetry group is non compact
$[SL(2,\R ) ]^n $, where $n$ is the number of moduli, and in particular
this gives flat directions in the scalar potential. They can, due to
their symmetry, eventually be considered as the point-field limit of
compactified superstring models.

In fact, it is possible to construct models such that the constraint is
exactly in the Veltman form, if we abandon the duality symmetries. A
general expression for a Veltman type condition can be easily obtained
for the generic class of models defined in eq.(\ref{eq:ar3})  using
eq.(\ref{eq:ar5}). The result is
\bea
 \hat \lambda_{i_A i_B i_C} \ \hat
\lambda^{i_A i_B i_C+} = e^{K_0} ({K_A}^{-1})^{i_A}_{j_A}
({K_B}^{-1})^{i_B}_{j_B} ({K_C}^{-1})^{i_C}_{j_C} \
\lambda_{i_A i_B i_C} \ \lambda^{j_A j_B j_C+} \
. \label {eq:ar33}
\eea
 Using eq.(\ref{eq:ar12}-\ref{eq:ar14}) we can
explicitely check that such a constraint violates the duality
symmetries, as proved more generally in eq.(\ref{eq:ar15} -
\ref{eq:ar16}).

It is nonetheless easy to construct models with the remnant symmetry
$U(1)^n \times $ diagonal \ scale symmetry in the K\"ahler potential.
The transformations of the moduli are
\bea
 \left \{ \bal T_\alpha
\rightarrow T_\alpha + i b_\alpha \\ \\ T_\alpha \rightarrow a T_\alpha
\ . \label{eq:ar34} \eal \right.
\eea
 They are sufficient in order to
get the flat directions and to forbid renormalizable terms in the
superpotential $W$. The global scale invariance is characteristic of
string models in the limit of large moduli.

A first example has no Higgs field and the superpotential is given in
eq.(\ref{eq:ar22}). The required K\"ahler potential is
\bea
 K = -{3
\over n} \sum_i \ell n \left [ t_i - \left ( {t_i^2 \sum_{j=1}^n t_j
\over \prod_{k=1}^n t_k^{3 \over n} } \right )^{1 \over 3} |\phi_i |^2
\right ] - \ell n (S + S^+) \ . \label{eq:ar35}
\eea
 Putting $\lambda_i
= \lambda$ for simplicity, the constraint is
\bea
 \sum_{i=1}^n \hat
\lambda_i^2 = {\lambda^2 \over s + s^+} \label{eq:ar36}
\eea
 and it is
of the Veltman type, eq.(\ref{eq:ar1}).

A second example contains a Higgs field $H$ plus $n$ quark fields
$\phi_i$. The model is defined by
\bea
 \bal K& = -{3 \over n}
\sum_{i=1}^n \ell n \left ( t_i - |\phi_i |^2 - ({t_i^3 / \prod_{j=1}^n
t_j^{3 \over n}}) | H |^2 \right ) - \ell n (S + S^+) \ , \\ W& = {1
\over 2} (\sum_i \lambda_i \phi_i^2 ) H \ .  \eal\label{eq:ar37}
\eea

The resulting constraint is the same as in eq.(\ref{eq:ar36}). We
consider that the last two models are less interesting in that they have
less symmetries and cannot be considered as effective orbifold models
with spontaneous supersymmetry breaking, as in the previous examples. In the
following section we will concentrate on the
multiplicative constraints.

Some remarks should be made about the induced soft supersymmetry
breaking terms. In the so-called large hierarchy compatible
supergravities \cite{FKZ}
all of them depend on the gravitino mass and on the
scaling weight of the metric for the chiral fields. They are independent
of the Yukawas and should be treated as dynamical variables if one
minimizes with respect to $m_{3/2}$.  In the gaugino condensation
scenario, at  tree level none of these terms appear in the observable
sector. At the one loop level, gaugino masses are induced \cite{BDGH} and
they depend on the Planck and gravitino mass, but not on Yukawas.
Through gauge interactions, they produce in principle all the other
breaking terms, which in the lowest order will not depend on Yukawas. In
the next paragraphs we will always consider the soft terms as being
independent of the Yukawas.

In order to simplify as much as possible the analysis, we suppose that
only the Yukawas of the quark type fields are dynamical and the others
are fixed. This is equivalent to fixing  some moduli fields which do not
contain the quark fields in their no-scale structure. For example, in
the model defined in eq.(\ref{eq:ar26}), fixing the moduli $T_n$ which
contains the Higgs field $H$ in the no-scale structure is equivalent to
fixing the coupling $\lambda_n$. In the following we will suppose, for
simplicity reasons, that the moduli related to the Higgs type fields
and, consequently, the Higgs Yukawa self-interactions, are fixed.  The
essential difference between the quark type Yukawas and the Higgs
self-interactions is that the second ones appear in the vacuum energy already
at tree level, for multiplicative type constraints. The first ones
contribute only at one loop  and
the generated hierarchy is radiatively induced, if they are
considered as dynamical variables.


\section{Constraints between Yukawa couplings and fermion mass
hierarchies.} \label{sec:nom3}

We start by clarifying the minimization with respect
to the Yukawas, in connection with the moduli
determination of the effective superstring theory.

Clearly, for our procedure to make sense, we must assume that some
moduli fields remain undetermined down to low energies, {\em i.e.}
down to energies below the supersymmetry breaking scale. Then the
orientation of the vacuum in the corresponding flat directions of the
potential will take place according
to the details of the low energy theory.
This, rephrased in the language of the low energy Yukawa couplings,
will lead to the dynamical determination of some of these Yukawa
couplings. Denoting by ${V_0} (\hat \lambda_i, \phi, m_{3/2})$ the
scalar potential in the observable sector, we have to minimize it
with respect to the yet undetermined moduli fields:
\bea
\sum_{I=1}^M
{\partial \hat \lambda_I \over \partial T_{\alpha}}
{\partial V_0 \over \partial \hat \lambda_I} = 0 \ . \label{eq:ar93}
\eea
In the presence of the constraints, the matrix ${\partial \hat \lambda_I
\over \partial T_{\alpha}}$ is degenerate, as expressed in
eq.(\ref{eq:ar91}) so the minimization with respect to the Yukawas is
subject to constraints as well. Obviously, the minimization should be
performed taking into account the constraints (\ref{eq:ar500}).

Using the RG invariance of $V_0$, we can write eq.(\ref{eq:ar93}) as a
function of the low energy Yukawas $\hat \lambda_I (\mu_0)$
\bea
{\partial \hat \lambda_I (M_P) \over \partial T_{\alpha}}
{\partial \hat \lambda_J (\mu_0) \over \partial \hat \lambda_I (M_P)}
{\partial V_0 \over \partial \hat \lambda_J (\mu_0)} = 0 \ .
\label{eq:ar94}
\eea
This equation has two solutions :

i) ${\partial \hat \lambda_J (\mu_0) \over \partial \hat \lambda_I (M_P)}
 = 0$, for any $I,J$. This may happen if $\hat \lambda_I (M_P)
\rightarrow \infty$, in which case all the Yukawas  reach their
maximally allowed values at $\mu_0$. This is the approach followed in
\cite{KPZ}, for example.

ii) ${\partial \hat \lambda_J (\mu_0) \over \partial \hat \lambda_I (M_P)}
 \not= 0$, $rank ( {\partial \hat \lambda_I (M_P) \over \partial
T_{\alpha}}
{\partial \hat \lambda_J (\mu_0) \over \partial \hat \lambda_I (M_P)}
 ) = M - p$.

In this case, minimizing with respect to
the moduli is equivalent to minimizing $V_0$ with respect to the
Yukawas at $\mu_0$, using as constraints
\bea
F_i \left(\hat \lambda (M_P) [\hat \lambda (\mu_0)]\right) = C_i \ . \
\eea
This solution is more interesting for generating the mass hierarchy
between the fermions and corresponds generally to the Nambu mechanism
discussed in the Introduction. The constraints $F_i$ are generically
more complicated than the simple eq.(1), but the qualitative results
are similar.

We will work under the hypothesis that the second solution ii) corresponds
to the real vacuum and that a non-trivial minimization must be performed
at low energy $\mu_0$.

An additive constraint of the
Veltman type (\ref{eq:ar1}) was analyzed in \cite{Nam} and it was shown
to produce a hierarchy between the fermion masses, irrespective of the
value of the constant $a$.

We  now analyze in detail the multiplicative constraints of the type
(\ref{eq:ar24}). If we are interested in the effective spontaneously
broken supersymmetric theory at a scale $\mu_0$, we must run
eq.(\ref{eq:ar24}) from $M_P$ to $\mu_0$ using the renormalization group
(RG) equations for the effective renormalizable theory.

To compute the vacuum energy at the low-energy scale $\mu_0 \sim
M_{susy}$ we proceed in the usual way. Using boundary values for the
independent model parameters at the Planck scale $M_P$ (identified here
with the unification scale), we evolve the running parameters down to
the scale $\mu_0$ using the RG equations and use the effective potential
approach \cite{CW}. The one-loop effective potential has two pieces
\bea
V_1 (\mu_0) = V_0 (\mu_0) + \Delta V_1 (\mu_0) \ , \label{eq:ar101}
\eea
where $V_0 (\mu_0)$ is the renormalization group improved tree-level
potential and $\Delta V_1 (\mu_0)$ summarizes the quantum corrections
given by the formula
\bea
\Delta V_1 (\mu_0) = {(1 / 64 \pi^2)} \ Str M^4 \ (\ln {M^2 \over \mu_0^2} -
{3 \over 2}) \ . \label{eq:ar102}
\eea
In (\ref{eq:ar102}) $M$ is the field-dependent mass matrix and all the
parameters are computed at the scale $\mu_0$. The vacuum state is
determined by the equation $\partial V_1 / \partial \phi_i = 0$, where
$\phi_i$ denotes collectively all the fields of the theory. The vacuum
energy is simply the value of the effective potential computed at the
minimum.
Let us define an 'average' mass $\bar m$.
\footnote{ Writing the decomposition
$Str M^4 \ (\ln M^2 /
\mu_0^2 - 3/2) = Str M^4 \ (\ln {\bar m}^2 /
\mu_0^2 - 3/2) + Str M^4 \ \ln M^2 / {\bar m}^2$, one can choose
$\bar m$ such that the second term in the decomposition,
containing logarithmic terms in the Yukawas, has a minimal contribution.}
In the following
the minimization process is always performed for $\mu_0^2 > {\bar m}^2
e^{-3/2}$.

For the toy model with no Higgs, eq.(\ref{eq:ar17}), the RG equations for
$\lambda_1$ and $\lambda_2$ are completely decoupled and can be easily
integrated. The constraint (\ref{eq:ar21}) can then be rewritten in the
form (two couplings)
\bea
 {\lambda^2_1 (\mu_0 ) \over 1 - {3
\lambda_1^2(\mu_0 ) \over 16 \pi^2 } \ln {M_P \over \mu_0}}\ \
 \ {\lambda^2_2 ( \mu_0 ) \over 1 - {3 \lambda^2_2 (\mu_0 ) \over 16
\pi^2 } \ln {M_P \over \mu_0 }} = \widetilde a ^4 \ , \label {eq:ar38}
\eea
 where $\lambda_i ( \mu_0 )$ are the effective couplings at the scale
$\mu_0$  and we omitted the hat notation for simplicity. If we were in
a perturbative regime ${3 \lambda_i^2 (\mu_0 ) \over 16 \pi^2} \ln {M_P
\over \mu_0} << 1$, an expansion of eq.(\ref{eq:ar38}) would give us
a Veltman-type constraint (\ref{eq:ar1}) with $a^2 = {16 \pi^2 \over 3
\ln {M_P \over \mu_0}}$.

In order to have the Nambu mechanism, we will suppose that at a
low-energy $\mu_0$ the fields $\phi_i$ decouple and can be fixed at
their minimum, their masses being larger than that of the moduli (for
realistic models such as the MSSM, or more generally models with Higgs
fields, this hypothesis is not necessary).
The vacuum energy at a low energy scale $\mu_0$ for a theory with soft
scalar masses $\mu$  is given by \cite{BD}
\bea
 {\cal E}_0 \ (\mu_0) = - M^4 \ [\xi_1 (1 - x_1 ) + \xi_2
(1 - x_2)] \ (\mu_0) \ , \label{eq:ar39}
\eea
 where we neglected the logarithmic terms in the
Yukawas and we used the notations
\bea
 x_i = {\lambda_i^2 \over 8\pi^2} \ln {\tilde \mu_0^2} \ , \ \xi_i = {1 \over
x_i} \left ( {1 - 2 x_i \over 1 - x_i}
\right ) ^2 , \ i = 1,2 \label{eq:ar40}
\eea
 and $ M^4 = {\mu^4 \over 32 \pi^2} \ln {\tilde \mu_0^2}$, with
$\tilde \mu_0^2 = \mu_0^2 e^{3/2} / {\bar m}^2$.

The lowest-energy configuration for (\ref{eq:ar38}) and (\ref{eq:ar39})
is obtained for $\xi_1 \rightarrow \infty$  ($x_1 \rightarrow 0$)
 and $\lambda^2_2 (\mu_0 ) = {16 \pi^2
\over 3 \log {M_p \over \mu_0}}$ or the  solution obtained by
exchanging $\lambda_1 \leftrightarrow \lambda_2$.

In this approximation one fermion remains massless whereas the other
one becomes massive. Its corresponding Yukawa coupling reaches the
``triviality bound'' \cite{CMPP}, characterized by $\lambda^2_2(M_P)
>> 1$. We are thus not in a perturbative regime for $\lambda_2$ and we
cannot develop in a series eq.(\ref{eq:ar38}). Even in the nonlinear
form,
however, eq.(\ref{eq:ar38}) produces the same mechanism of generating
hierarchies. The difference with respect to the Veltman-type constraint
(\ref{eq:ar1}) is that the mass of the massive fermion is now controlled
by the triviality bound. Taking $n$ Yukawa couplings constrained by
eq.(\ref{eq:ar24}) will always give $n-1$ massless fermions and a
massive one.

One can relate the determination of the couplings (or, equivalently,
the moduli) to the spontaneous breakdown of the residual diagonal
dilatation symmetry (\ref{eq:ar34}), which acts on the fields as
$\phi_i \rightarrow a^{1/2} \phi_i$. Soft mass terms in the
tree-level potential induce nonzero vev's $<\phi_i>$ which break the
dilatation symmetry. Supersymmetry breaking is, consequently, an
important ingredient in the mechanism. Remark that in the absence of the
constraint (\ref{eq:ar38}), the minimization process would force both
couplings $x_1$, $x_2$ to vanish thus forbidding any hierarchy between
the fermion masses.

We now turn to a more realistic case containing Higgs fields and
described by eq.(\ref{eq:ar26}) with $n = 3$ and the multiplicative
constraint (\ref{eq:ar28}). It is for example the case of the MSSM
when one considers
only two u-type quarks and it is the ideal example for understanding the
hierarchies between the fermion generations. In order to have complete
analytical expressions, we will limit ourselves to the case
$\lambda_3 = 0$, which reproduces the essential features of the
mechanism.

Using the effective potential formalism and neglecting the logarithmic
corrections we can write the vacuum energy (the dependence in
$\lambda_1$, $\lambda_2$ ) as
\bea
 {\cal E}_0 = - A(\lambda^2_1 +
\lambda^2_2 ) \label{eq:ar41}
\eea
 with $A > 0$. To obtain this, we add
the most general soft breaking terms at low-energy $\mu_0$ and compute
$Str \ M^4$, where
\bea
 \mbox{Str} {M}^n = \sum_J (-1)^{2J} (2J + 1) Tr
M_J^n \ . \label{eq:ar42}
\eea
 The quark-type fields $\phi_i$ have zero
vev's and the Higgs vev's $<H_i>$ are held fixed and independent of
$\lambda_i$'s. The expression (\ref{eq:ar41}) is typical of
the MSSM, which
will be studied in detail in the next paragraph.  The first step is to
run the constraint (\ref{eq:ar28}) for $n = 3$ from $M_P$ to $\mu_0$
using the RG equations
\bea
 \left \{ \bal \mu \ {d \over d \mu} \lambda_1 =
{\lambda_1 \over 32 \pi^2} (5 \lambda^2_1 + \lambda^2_2 ) \\ \\ \mu \ {d
\over d \mu} \lambda_2 = {\lambda_2 \over 32 \pi^2} (5 \lambda^2_2 +
\lambda_1^2) \ . \label{eq:ar43} \eal \right.
\eea
 For arbitrary
initial values $\lambda_i(M_P)$ that fulfill eq.(\ref{eq:ar28}) this
cannot be done analytically. Exact integration of (\ref{eq:ar43}) is
possible if $\lambda_1(M_P) = \lambda_2(M_P)$ and an approximate
solution is found if $\lambda_2(M_P) << \lambda_1(M_P)$. Obviously, the
corresponding Yukawas at low energy $\mu_0$ will be equal
$\lambda_1(\mu_0 ) = \lambda_2(\mu_0)$ in the first case and will
satisfy $\lambda_2(\mu_0 ) << \lambda_1(\mu_0 )$ in the second. We are
then able to compare the vacuum energy for the two configurations and to
decide whether the minimum corresponds to a symmetric or an
asymmetric solution. A complete numerical analysis will follow for the
whole set of boundary conditions $\lambda_i(M_P)$.

i) $\lambda_1(M_P) = \lambda_2(M_P)$.  Due to the symmetry of
eq.(\ref{eq:ar43}) we will have $\lambda_1(\mu_0) = \lambda_2(\mu_0)
\equiv
\lambda(\mu_0 )$. A straightforward integration of eq.(\ref{eq:ar43})
gives
\bea
 \lambda^2(M_P) = {\lambda^2(\mu_0) \over 1 - {3\lambda^2(\mu_0)
\over 8\pi^2} \ln {M_P \over \mu_0}} = \widetilde a ^2 \ .
\label{eq:ar44}
\eea
where $\widetilde a^2 = \lambda^2 / s$.
 The vacuum energy has the expression
\bea
 {\cal
E}_0(i) = - 2 A \lambda^2(\mu_0) = - 16\pi^2 A {\widetilde a^2 \over 8
\pi^2 + 3 \widetilde a^2 \ln {M_P \over \mu_0}} \label{eq:ar45}
\eea

ii) $\lambda_2(M_P) << \lambda_1 (M_P)$.  In this case
eq.(\ref{eq:ar43}) can be approximately integrated, and the solution is

\bea
 \left \{ \bal \lambda^2_1(\mu_0 ) = {\lambda^2_1(M_P) \over 1 + {5
\lambda_1^2 (M_P) \over 16\pi^2} \ln {M_P \over \mu_0} }\\ \\
\lambda^2_2(\mu_0 ) = {\lambda^2_2(M_p) \over \left[
 1 + {5 \lambda_1^2 (M_P) \over 16\pi^2} \ln {M_P \over \mu_0}
\right]^{1 \over 5}} \ .
\label{eq:ar46} \eal\right.
\eea

The constraint eq.(\ref{eq:ar28}) in the case $\lambda^2_2(\mu_0 )
\simeq 0$ will then impose

\bea
 \left \{ \bal \lambda^2_2(\mu_0 ) \simeq 0\\ \\ \lambda^2_1(\mu_0 )
\simeq {16 \pi^2 \over 5 \ell n {M_p \over \mu_0}} \ . \label{eq:ar47}
\eal\right.
\eea
 $\lambda^2_1(\mu_0 )$ is determined, as in the toy
model, by the triviality bound, corresponding to $\lambda^2_1(M_p) >>
1$. We will pospone the question whether the small
couplings are exactly zero or not to the next section. There a detailed
analysis in the context of the MSSM will show that they indeed are
different from zero and a fermion mass hierarchy is generated. This is
not relevant for the present analysis as long as the small coupling
is negligible compared to the large one, as should be clear from the
additive form of the ${\cal E}_0$, eq.(\ref{eq:ar41}).

The vacuum energy is approximately given by
\bea
 {\cal E}_0(ii ) =
-{16 \pi^2 A \over 5 \ell n {M_p \over \mu_0}} \ . \label{eq:ar48}
\eea

Comparing the two energies (\ref{eq:ar45}) and (\ref{eq:ar48}) we
obtain the condition needed in order to have
${\cal E}_0(ii) < {\cal E}_0(i)$.
Explicitly we have $\widetilde a^2 < {4 \pi^2 / \ln {M_P \over
\mu_0}}$. Substituting $\widetilde a^2$ with the expectation value
of the dilaton $s$, we get

\bea
 s  > {\lambda^2 \ln {M_P \over \mu_0 } \over 4 \pi^2} \ .
\label {eq:ar49}
\eea
 This is the main difference between a
multiplicative-type constraint and an additive type constraint. In the
first case, a minimal value for the vev of dilation $s$ is necessary in
order to have the Nambu mechanism, whereas in the second case no
condition is needed.

It should be remarked that the RG equations (\ref{eq:ar43}) have an infrared
fixed
point ${\lambda_1^2 / \lambda^2_2} = 1$ which corresponds to the
symmetric case $(i)$ above. The previous analysis tells us that, if
eq.(\ref{eq:ar49}) is satisfied, this solution is disfavored compared
with the asymmetric one $\lambda_2 << \lambda_1$ and the infrared fixed
point is not reached.
A detailed numerical analysis shows that the solution $(ii)$ is the
absolute minimum.

Adding a third quark coupling does not change qualitatively the results.
The prefered configuration has two very small Yukawas and the third one
fixed by the triviality bound. The only change is in eq.(\ref{eq:ar49})
where the factor $4\pi^2$ should be replaced by $2\pi^2$. The system
will never reach the infrared fixed point ${\lambda^2_2 / \lambda^2_1} =
{\lambda^2_3 / \lambda^2_1} = 1$.

\section{The case of the Minimal Supersymmetric Standard Model (MSSM).}
\label{sec:nom4}

We now analyse  the MSSM model \cite{Fay} and the possible
phenomenological constraints which must be satisfied in order
for the Nambu mechanism to work.
We first concentrate  on the sign of the $A$ coefficient of
 eq.(\ref{eq:ar2}), in the leading \ $\ln \Lambda^2$
approximation and try to be as general as possible regarding the
constraint between Yukawas. As long as the constant $\widetilde a^2$ in
eq.(\ref{eq:ar21}) is sufficiently small, eq.(\ref{eq:ar49}), the
positivity of $A$ is the signal of a generation of hierarchies, both for
additive and multiplicative  constraints. We then compute the bottom
coupling in the MSSM, neglecting the Yukawas of the first two generations
and using the multiplicative constraint eq.(\ref{eq:ar21}). The value
obtained is naturally small compared with the top mass, although we do
not obtain an exponential hierarchy of the Nambu type (~which would
anyway be too large to account for the observed ratio of masses~).

The superpotential for the MSSM is given by the formula
\bea
 W = \lambda_U^{ij} Q^i
U^{cj} H_2 + \lambda_D^{ij} Q^i D^{cj} H_1 + \lambda_L^{ij} L^i E^{cj}
H_1 + \mu H_1 H_2
\label{eq:ar50}
\eea
where $i,j=1,2,3$ are generation indices and the soft-breaking terms read
\bea
 \bal
-{\cal L}_{soft} = M^2 (|z_Q|^2 + |z_{U^c}|^2 + |z_{D^c}|^2 ) +
\sum_{k=L,E^C} M_k^2 |z_k|^2 + m_1^2 |z_1 |^2 + m_2^2 |z_2|^2 \\ \\ +
m^2_3 (z_1 z_2 + z_1^+ z_2^+ ) + ({\cal A}_U^{ij} z_U^i z_{U^c}^j z_2 +
{\cal A_D}^{ij} z_D^i z_{D^c}^j z_1 + {\cal A}_L^{ij} z_L^i z_{L^c}^j
z_1 + h.c. ) +\\ \\ + {M_3 \over 2} (\lambda^A_3
\lambda^A_3 + \bar \lambda^A_3 \bar \lambda^A_3 ) + {M_2 \over 2}
(\lambda^i_2 \lambda^i_2 + \bar \lambda^i_2 \bar \lambda^i_2 ) + {M_1 \over 2}
(\lambda_1
\lambda_1 + \bar \lambda_1 \bar \lambda_1 ) \ , \label{eq:ar51} \eal
\eea
where $z_x$ is the scalar component of the chiral superfield $X$, $z_i$
($i=1,2$)
are the two Higgs fields and $\lambda_3^A$, $\lambda_2^i$, $\lambda_1$ are the
$SU(3)$, $SU(2)$, $U(1)$ gaugino fields.
 Keeping only the neutral scalar fields for $H_1$ and $H_2 , z_1$
and $z_2$ of vev's $v_1$ and $v_2$, the tree-level scalar potential reads

\bea
 V_0 = (\mu^2 + m^2_1) |z_1|^2 + (\mu^2 + m^2_2 ) |z_2|^2 + m^2_3
(z_1 z_2 + z_1^+ z_2^+ ) + {g^2_1 + g_2^2 \over 8} (|z_1|^2 -
|z_2|^2 )^2 \ , \label{eq:ar52}
\eea
 where $g_1$ and $g_2$ are the $U(1)$
and $SU(2)$ coupling constants, respectively.  An important parameter of
the theory is the angle $\beta$ defined by $tg \beta = {v_2 / v_1}$,
expressed after minimization of $V_0$ in terms of the other parameters by
\bea
 sin \ 2\beta = {- 2m^2_3
\over 2 \mu^2 + m^2_1 + m^2_2} \ . \label{eq:ar53}
\eea
 As expected,
there is no Yukawa dependence of the vacuum energy at tree level,
eq.(\ref{eq:ar52}) and we must go to the one-loop level. In the leading
\ $\ln \Lambda^2$ approximation, the vacuum energy is determined by
$Str M^4$ which reads
\bea
 \bal { 1 \over 3} \ Str M^4 = \left [
4(\mu^2/ tg^2 \beta + 2M^2) - (g^2_1 + g^2_2 ) (v^2_2 - v^2_1) \right
] Tr \lambda^2 _{U} v^2_2 + \\ \\ + \left [ 4(\mu^2 tg^2 \beta +
2M^2) + (g^2_1 + g^2_2 ) (v^2_2 - v^2_1) \right ] Tr (\lambda^2
_{D} + {1 \over 3} \lambda_{L}^2 ) v^2_1 + \\ \\ 8 \mu Tr (\lambda_U {\cal
 A_U} + \lambda_D {\cal A_D} + {1 \over 3} \lambda_L {\cal A_L}) v_1 v_2
\\ \\ \equiv A_U Tr \lambda^2_{U} +
A_D Tr (\lambda^2_{D} + {1 \over 3} \lambda^2_{L}) + 8 \mu Tr (\lambda_U
{\cal
 A_U} + \lambda_D {\cal A_D} + {1 \over 3} \lambda_L {\cal A_L}) v_1 v_2 \ .
\label{eq:ar54} \eal
\eea
 The last
line defines the parameters $A_U$ and $A_D$. Using the $Z$ mass
expression $M^2_Z = {1 \over 2} (g^2_1 + g^2_2 ) (v^2_1 + v^2_2 ) $, we
can rewrite $A_U$ and $A_D$ as
\bea
\bal A_U = 2 \ [ 2\mu^2 / tg^2 \beta +
4 M^2 - M^2_Z + (g^2_1 + g^2_2) v^2_1 ] \ v^2_2 \ , \\ \\ A_D = 2 \ [ 2\mu^2 \
tg^2 \beta + 4 M^2 - M^2_Z + (g^2_1 + g^2_2) v^2_2 ] \ v^2_1 \ .
\label{eq:ar55} \eal
\eea
 In order to decide about the signs of $A_U$
and $A_D$ we use the experimental inequality
\bea
 (Str
M^2)_{\mbox{quarks + squarks}} = 4M^2 > M^2_Z \ . \label{eq:ar56}
\eea

Then $A_U , A_D > 0$ and the vacuum energy in the leading $\log
\Lambda^2$ approximation has the Nambu form, eq.(\ref{eq:ar2}), with
$B=0$ and an additional linear term which does not change the shape
of the vacuum energy as a function of the Yukawas, but will play an
essential role in the minimisation process.

The positivity of $A_U$, $A_D$ is a direct consequence of supersymmetry
and is due to the Yukawa dependent bosonic contributions in
(\ref{eq:ar54}). In the non-supersymmetric Standard Model the sign
is negative and the present considerations would not apply.
Using eq.(\ref{eq:ar101}) and eq.(\ref{eq:ar102}), we obtain the vacuum
energy as a function of $\lambda_U$ and $\lambda_D$, which is a
paraboloid unbounded from below. If we had no constraint, both Yukawas
would tend to the maximally allowed values and no hierarchy
would be generated. The role of the constraint, as emphasized in
\cite{BD} is to restrict the coupling constant parameter space so that the
minimum of the vacuum energy is {\it exactly} where the mass hierarchy
occurs.

Because the Nambu factor $A_D$ for the
D-type quarks is three times larger than the corresponding one for the
leptons, the heaviest fermion obtained by minimization is always
a quark and not a lepton.
In order to decide whether the heavy quark will be of the $U$ or of the $D$
type, we must compare $A_U$ and $A_D$. If $A_U > A_D$, a $U$ type quark
becomes massive and all the others remain light (in the
leading \ $\log \Lambda^2$ approximation). Using the definitions in
eq.(\ref{eq:ar54}), we find
\bea
 A_U - A_D = 2 (tg^2\beta - 1) (- 2\mu^2
+ 4M^2 - M^2_Z ) v^2_1 \ . \label{eq:ar57}
\eea

 The region in the parameter space where $A_U > A_D$ is described by
the inequality:
\bea
tg^2 \beta > {2 M^2 + m_1^2 \over 2 M^2 + m_2^2 } \ . \label{eq:ar58}
\eea
We therefore need a minimal critical value for $tg \beta$ of order one,
which depends
on the soft masses, in order to have a heavy top quark.
As we will explicitely
check, and in a way similar to the toy model considered in the preceding
paragraph,  the Nambu mechanism is
dictated by the sign of $A_U$ and $A_D$ and a value of
the dilaton vev larger than a critical value. Then eq.(\ref{eq:ar58}) states
that there is no need of fine tuning in order to understand the hierarchy
between
the top quark and the other fermions.

In what follows, the Yukawas for the first two generations will be
neglected; only the top and the bottom Yukawas, denoted $\lambda_U$ and
$\lambda_D$ will be considered.

In order to do the low energy minimization of the vacuum energy, we will
proceed in two steps. First of all, we will show that a non-trivial minimum
for $\lambda_D$ appears in the MSSM compatible with $\lambda_D / \lambda_U
<<1$. An analytic
expression will be derived, as function of the MSSM parameters. A second step,
as above, is the comparison of the energy of this extremum
with that of the symmetric solution $\lambda_D = \lambda_U$.

The RG equations in the case $\lambda_U (M_P) = \lambda_D(M_P)$
simplify, because in this case $\lambda_U(\mu_0 ) = \lambda_D(\mu_0 )
\equiv \lambda(\mu_0)$ for any $\mu_0 < M_P$. It reads ($g_3$ and $g_2$
are the $SU(3) \times SU(2)$ coupling constants)
\bea
 \mu {d \over d
\mu} \lambda = {\lambda \over 16\pi^2}\ (7\lambda^2 - {16 \over 3} g^2_3
- 3g^2_2 ) \ . \label{eq:ar59}
\eea
 Defining \cite{FL}
\bea
 \gamma^2
(Q) = e^{-{ 1 \over 8\pi^2} \int^Q_{M_P} ({16 \over 3} g_3^2 + 3 g^2_2
) dt} \ , \label{eq:ar60}
\eea
 the solution of eq.(\ref{eq:ar59}) is
given by
\bea
 \lambda^2(\mu_0 ) = {\gamma^2 (\mu_0 )
 } \ \ \ {\lambda^2(M_P) \over 1 + {7 \lambda^2(M_P) \over 8 \pi^2
\gamma^2 (M_P)} \int_{\mu_0}^{M_P} \gamma^2(Q) d \ln Q } \ .
\label{eq:ar61}
\eea
 In the case $\lambda_U(M_P) >> \lambda_D(M_P)$, the RG equation are
approximately
given by
\bea
\bal
 \mu
{d \over d\mu} \lambda_U = {\lambda_U \over 16\pi^2} (6\lambda_U^2 - {16
\over 3} g_3^2 - 3g_2^2 ) \\ \\
{d \over d\mu} \lambda_D = {\lambda_D \over 16\pi^2} (\lambda_U^2 - {16
\over 3} g_3^2 - 3g_2^2 )
\label{eq:ar62}
\eal
\eea
 and, for $\lambda_U(M_P)
>> 1 $ , an approximate solution of (\ref{eq:ar62}) is
\bea
\bal
  \lambda_U^2(M_P ) = {1 \over \gamma^2 (\mu_0 ) } \ \ \ {\lambda_U^2(\mu_0)
\over 1 - {6 \lambda_U^2(\mu_0) \over 8 \pi^2
\gamma^2 (\mu_0)} \int_{\mu_0}^{M_P} \gamma^2(Q) d \ln Q } \ , \\ \\
\lambda_D (M_P) = {\lambda_D (\mu_0) \over \gamma_3 (\mu_0)} e^{(1 / 16 \pi^2)
\int_{\mu_0}^{M_P}
\lambda_U^2 d \ln Q}
 \ . \label {eq:ar63}
\eal
\eea
The constraint (\ref{eq:ar21})
\bea
\lambda_U (M_P) \lambda_D (M_P) = {\widetilde a}^2 \label{eq:ar150}
\eea
in this approximation reads explicitly
\bea
 {1 \over \gamma^4 (\mu_0 ) } \ \ \ {\lambda_U^2(\mu_0) \lambda_D^2 (\mu_0)
\over 1 - {6 \lambda^2(\mu_0) \over 8 \pi^2
\gamma^2 (\mu_0)} \int_{\mu_0}^{M_p} \gamma^2(Q) d \ell n Q } e^{(1 / 8 \pi^2)
\int_{\mu_0}^{M_P}
\lambda_U^2 d \ln Q } = \widetilde a^4 \ . \label{eq:ar151}
\eea
It is highly non-linear and difficult to deal analytically with . We
make the hypothesis that the minimum lies close to the top
quasi-infrared fixed point and linearize around this point.
A self-consistency check will be performed at the end.
First of all, we trade $\lambda_U$ in favor of a new variable $\delta$
defined as
\bea
\delta^2 = 1 - {6 \lambda^2(\mu_0) \over 8 \pi^2
\gamma^2 (\mu_0)} \int_{\mu_0}^{M_P} \gamma^2(Q) d \ln Q  \ , \label{eq:ar152}
\eea
such that
\bea
\lambda_U^2 (\mu_0) =  {8 \pi^2
\gamma^2 (\mu_0) \over 6 \int_{\mu_0}^{M_P} \gamma^2(Q) d \ln Q } (1 -
\delta^2) =
x_0^2 (1 - \delta^2) \ . \label{eq:ar153}
\eea

The constraint (\ref{eq:ar151}) in the limit $\delta <<1$ gives us
\bea
\lambda_D^2 (\mu_0) = c^2 {\delta^2 / x_0^2} \ , \label{eq:ar154}
\eea
where
\bea
c^2 = \widetilde a^4 \gamma^4 (\mu_0) e^{- (1 / 8 \pi^2) \int_{\mu_0}^{M_P}
x_0^2 d \ln Q }
\ . \label{eq:ar155}
\eea

Thanks to the constraint, we only have one variable, $\delta$, to be minimized
in the effective potential. The solution $\lambda_D = 0$ corresponds to
$\delta = 0$, so we are interested in the small $\delta$ limit.

The one-loop effective potential, neglecting the logarithmic corrections
(we will come back later to discuss their effect) can be cast in the simple
form
\bea
V_1 (\mu_0) = V_0 (\mu_0) - ({A'_U} \lambda_U^2 + {A'_D} \lambda_D^2) -
\alpha ({\cal A_U} \lambda_U + {\cal A_D} \lambda_D) \ , \label{eq:ar156}
\eea

where we defined the functions
\bea
\bal
{A'}_{U,D} = {(3 \ln {\tilde \mu_0}^2 / 64 \pi^2)} A_{U,D} \\ \\
\alpha = (24 \mu / 64 \pi^2) \ln {\tilde \mu_0}^2 (\lambda_U {\cal A_U} +
\lambda_D {\cal A_D}) v_1 v_2 \ , \label{eq:ar157}
\eal
\eea
$\tilde \mu_0$ being defined as in eq.(\ref{eq:ar40}).

Using the explicit expressions for $\lambda_{U,D}$ (\ref{eq:ar153})and
(\ref{eq:ar154}),
$V_1$ appears simply as a function at most quadratic in $\delta$
\bea
V_1 (\mu_0) = cst + ({A'_U} x_0^2 - {{A'_D} c^2 \over x_0^2} + {\alpha \over2}
{\cal A_U} x_0) \delta^2 - \alpha {{\cal A_D} c \over x_0} \delta \ .
\label{eq:ar158}
\eea

We minimize this expression with respect to $\delta$, keeping all the other
parameters fixed, particularly the Higgs vev's $v_1$ and $v_2$. Two
possibilities
arise:

$i)$ If
\bea
{\widetilde a^4} < \left [ x_0^3 (2 {A'_U} x_0 + \alpha {\cal A_U}) / 2 {A'_D}
\gamma^4
(\mu_0) \right ] e^{(1 / 8 \pi^2) \int_{\mu_0}^{M_P} x_0^2 \ln Q}
\label{eq:ar159}
\eea

we have a minimum for $\delta$, given by
\bea
\delta = \alpha {\cal A_D} c / (2 {A'_U} x_0^3 - {2 {A'_D} c^2 \over x_0} +
{\alpha}
{\cal A_U} x_0^2) \ . \label{eq:ar160}
\eea

$ii)$ If
\bea
{\widetilde a^4} > \left [ x_0^3 (2 {A'_U} x_0 + \alpha {\cal A_U}) / 2 {A'_D}
\gamma^4
(\mu_0) \right ] e^{(1 / 8 \pi^2) \int_{\mu_0}^{M_P} x_0^2 \ln Q}
\label{eq:ar161}
\eea
the extremum becomes a maximum.

Generically (for a large region of the parameter space) $\delta <<1$ if
$\widetilde a^2 < 1$. Taking into account that $\widetilde a^2$ is related to
the
gauge coupling constant value at the Planck scale (so to the dilaton
vev), this is a reasonable
assumption for weakly coupled effective string models. Consequently,
from now on we place ourselves in the case $i)$.

The ratio of the two Yukawa couplings at this extremum is given by
\bea
{\lambda_D / \lambda_U} =  {\widetilde a^4} \alpha {\cal A_D} \gamma^4 (\mu_0)
e^{- (1 / 8 \pi^2) \int_{\mu_0}^{M_P} x_0^2 d \ln Q } / (2 {A'_U} x_0^3 -
{2 {A'_D} c^2 \over x_0} + {\alpha}
{\cal A_U} x_0^2) \ . \label{eq:ar162}
\eea

A parameter space analysis of this relation shows that for a large region and
${\widetilde a^4} \sim 1/4$ (a phenomenologically reasonable value), $\lambda_D
/ \lambda_U <<1$ and {\cal no large value for $tan \beta$} is needed in order
to correctly reproduces the top and bottom masses from the experimental data.

Let us note that the $\mu$ parameter of MSSM \cite{Kim} is essential to produce
a non-vanishing value for $\lambda_D$.

We finally come back to the logarithmic corrections which were neglected in
$V_1 (\mu_0)$. These could be important and change qualitatively the
conclusions if they dominate in the small $\lambda_D$ limit. An explicit
computation of the
term $Str M^4 \ell n M^2$ shows that the relevant term behaves as $\lambda_D^4
 \ell n \lambda_D^2$ and is completely negligible compared to the linear term
considered above. So, compared to our original motivation, the toy model of
Nambu, the logarithmic corrections play no role in the MSSM and the bottom mass
is entirely due to a linear term proportional to the parameter $\mu$.

 We can now compute the
ratio of the two vacuum energies. The simplest case is
${\cal A_U} = {\cal A_D} = 0 $, in which case we obtain
\bea
 \bal \left | { {\cal
E}_0^{\lambda_U = \lambda_D} \over {\cal E}_0^{\lambda_U >> \lambda_D}}
\right | = {A_U + A_D \over A_U} \ {6 \lambda^2(M_P) \int_\mu^{M_P}
\gamma^2(Q) d \ln Q \over 8 \pi^2 \gamma^2 (M_P) + 7\lambda^2(M_P)
\int_\mu^{M_p} \gamma^2 (Q) d \ln Q  } \ . \label{eq:ar64} \eal
\eea
Using the fact that ${A_U + A_D \over A_U} < 2$, we find that a sufficient
condition for the configuration $\lambda_U >> \lambda_D$ to be
energetically preferred is
\bea
\lambda^2(M_P) < {8 \pi^2 \over 5} {\gamma^2 (M_P) \over \int^{M_P}_{\mu_0}
\gamma^2 (Q) d \ln Q} \ . \label{eq:ar65}
\eea
In the limit of very small gauge coupling constants, eq.(\ref{eq:ar65})
reduces to an inequality of the type (\ref{eq:ar49}) for the dilaton.
In the general case with $\cal A_U, \cal A_D$ different from zero, the
equivalent of the equation (\ref{eq:ar65}) becomes more involved,
but we always have an
upper bound for $\lambda^2(M_P)$ which is equivalent to a lower bound
for the dilaton $s$.
\section{Conclusions} \label{sec:nom5}

The aim of this paper is a dynamical understanding of the hierarchy
between the mass of the top quark and the other fermions. The central
hypothesis is that at the tree level of supergravity the theory has
flat directions which are lifted by the breaking of supersymmetry.
If the mass of the corresponding moduli is very small, of the order of
the electroweak scale, and if the low-energy Yukawa couplings are non-
trivial homogeneous functions of the moduli, the Yukawas can be
regarded as dynamical variables at low-energy. The homogeneity property
is natural in the context of the effective string models, which guarantee also
the existence of the flat directions. Our ignorance about the
supersymmetry breaking mechanism is hidden in the soft-breaking
parameters, which in turn will fix the couplings by the minimization
process. The procedure can be viewed also as a way to compute ratios of
the expectation values of the moduli fields ignoring the precise
mechanism of supersymmetry breaking.

The existence of constraints between the low-energy couplings is
automatic {\it if} the number of couplings which are {\it non-trivial}
functions of the moduli is greater or equal to the number of moduli.
The most interesting case is when the model has the same number of couplings
and moduli. In this case only one constraint is obtained at the Planck
scale, to be evolved down to the scale $\mu_0 \sim M_{susy}$.

Assuming K\"ahler type transformations for the K\"ahler potential under
the duality symmetries imposes a  multiplicative structure to the
constraints at the Planck scale. To obtain a Veltman-like additive constraint,
 we must give up the geometrical structure of the K\"ahler potential,
keeping only the axionic symmetries and a diagonal scale invariance for
the moduli. The multiplicative constraints generically put lower limits
on the dilaton in order for the mechanism to work, which in
superstring-inspired supergravity, allow us to consider only the
perturbative regime of the string.

When we apply this to the MSSM, we find that, due to supersymmetry, the
functional dependence of the vacuum energy is in a first approximation
as in the toy model of Nambu, eq.(\ref{eq:ar2}) with $A > 0$, $B =
0$. An additional linear term plays an important role in the
minimization process.
The mechanism predicts a heavy top quark if $tg \beta$ has a lower
limit of the order of one, given in eq.(\ref{eq:ar58}), and a heavy
bottom quark if this limit
is violated. The heavy fermion can never be a lepton due to the small
coefficient in front of its Yukawa coupling in the vacuum energy. A
lower limit on the dilaton vev must be imposed; its explicit value was
obtained in the symplifying case
of small gauge coupling constants and
vanishing trilinear soft-breaking terms. The condition
becomes more involved in the general case.

We computed analytically the bottom Yukawa coupling neglecting the Yukawas for
the first two generations, using the multiplicative constraint
(\ref{eq:ar150}) at the
Planck scale. The effective potential at a low scale $\mu_0 \sim M_Z$ is
minimized
with respect to $\lambda_U$ and $\lambda_D$, taking into account the
constraint translated at the scale
$\mu_0$ with the help of the RG equations. The top mass is very close to the
infrared effective fixed-point value, which is due essentially to the fact that
the minimization forces $\lambda_U (M_P) >>1$.
The presence in
the effective potential of a term linear in the Yukawas and proportional to the
$\mu$ parameter of MSSM turns out to
be essential for the stabilization of the bottom Yukawa to a small,
non-vanishing value.

The bottom mass is naturally small but not exponentially suppressed as in the
Nambu
example. The computed value is compatible with the existing data for a large
allowed region of the parameter space of the MSSM.

We insist on the fact that $tg \beta$ can be of order one and still the ratio
$m_b / m_t$ can easily to be made small, due to the small value of the
corresponding
Yukawas.

A complete phenomenological analysis would, of course, require the inclusion of
all the Yukawas. The difficult part of this program is probably in
extracting the
correct constraint(s) from the underlying string level. The large number of
soft-breaking terms can be substantially reduced by imposing some universal
boundary conditions at the Planck scale. In this way, the mechanism acquires
a predictive power and can be confronted with the known phenomenology,
all the Yukawas being dynamically determined.

\section{\bf Acknowledgements}
We would like to thank C. Kounnas for interesting discussions and
comments.

\end{document}